\documentclass[runningheads]{llncs}
\usepackage[T1]{fontenc}
\usepackage{graphicx,booktabs,booktabs,multirow,subfigure}
\usepackage[section]{placeins}
\usepackage{float}
\usepackage{orcidlink}
\usepackage{amsmath}
\usepackage{amssymb}
\usepackage{hyperref}
\makeatletter
\def\UrlAlphabet{%
      \do\a\do\b\do\c\do\d\do\e\do\f\do\g\do\h\do\i\do\j%
      \do\k\do\l\do\m\do\n\do\o\do\p\do\q\do\r\do\s\do\t%
      \do\u\do\v\do\w\do\x\do\y\do\z\do\A\do\B\do\C\do\D%
      \do\E\do\F\do\G\do\H\do\I\do\J\do\K\do\L\do\M\do\N%
      \do\O\do\P\do\Q\do\R\do\S\do\T\do\U\do\V\do\W\do\X%
      \do\Y\do\Z}
\def\UrlDigits{\do\1\do\2\do\3\do\4\do\5\do\6\do\7\do\8\do\9\do\0}
\g@addto@macro{\UrlBreaks}{\UrlOrds}%特殊符号
\g@addto@macro{\UrlBreaks}{\UrlAlphabet}
\g@addto@macro{\UrlBreaks}{\UrlDigits}
\makeatother

\begin{document}
\title{stEnTrans: Transformer-based deep learning for spatial transcriptomics enhancement}

\author{
Shuailin Xue$^1$ 
\and
Fangfang Zhu$^2$
\and
Changmiao Wang$^3$
\and
Wenwen Min$^{1*}$\orcidlink{0000-0002-2558-2911}
}
\authorrunning{Shuailin Xue et al.}
\institute{
$^1$School of Information Science and Engineering, Yunnan University, Kunming, China\\
$^2$College of Nursing Health Sciences, Yunnan Open University, Kunming, China\\
$^3$Shenzhen Research Institute of Big Data, Shenzhen, China\\
Correspondence author: minwenwen@ynu.edu.cn }
\maketitle  % typeset the header of the contribution
\begin{abstract}
The spatial location of cells within tissues and organs is crucial for the manifestation of their specific functions.Spatial transcriptomics technology enables comprehensive measurement of the gene expression patterns in tissues while retaining spatial information. However, current popular spatial transcriptomics techniques either have shallow sequencing depth or low resolution. We present stEnTrans, a deep learning method based on Transformer architecture that provides comprehensive predictions for gene expression in unmeasured areas or unexpectedly lost areas and enhances gene expression in original and inputed spots. Utilizing a self-supervised learning approach, stEnTrans establishes proxy tasks on gene expression profile without requiring additional data, mining intrinsic features of the tissues as supervisory information. We evaluate stEnTrans on six datasets and the results indicate superior performance in enhancing spots resolution and predicting gene expression in unmeasured areas compared to other deep learning and traditional interpolation methods. Additionally, Our method also can help the discovery of spatial patterns in Spatial Transcriptomics and enrich to more biologically significant pathways. Our source code is available at \url{https://github.com/shuailinxue/stEnTrans}.

\keywords{Spatial transcriptomics \and Deep learning  \and Self-supervised learning \and Transformer \and Imputation \and Prediction \and Enhancement}
\end{abstract}

\section{Introduction}
Traditional gene expression researches typically focus only on the overall gene transcript counts in entire tissues, lacking the preservation of spatial positioning information within the tissue. The spatial position of cells within tissues can have a significant impact on
their functions. Therefore, neglecting the spatial information of cells may lead
to an insufficient understanding of gene functions and regulatory mechanisms. Spatial transcriptomics (ST) technology aims to detect the quantity of gene transcripts within tissues while preserving spatial location information \cite{moses2022museum}. ST technology provides the capability to examine in detail the spatial distribution of gene expression at the tissue or cellular level. This allows researchers to gain a more accurate understanding of the gene expression patterns in different regions \cite{sun2020statistical,min2024dimensionality}, delving deeper into the exploration of cellular heterogeneity and interactions between neighboring cells. In light of the substantial advantages offered by ST technoloies, it have been used in various field of biology, such as tumor heterogeneity \cite{li2022spatial}, embryonic development \cite{lohoff2022integration} and Neuroanatomy \cite{huuki2023integrated}.

Existing ST technologies mainly fall into two categories: (1)imaging-based approaches,directly observing and quantitatively analyzing mRNA or protein expression in tissues without the need for prior RNA sequencing, such as STARmap \cite{wang2018three}. (2)next-generation sequencing (NGS)-based approaches \cite{asp2020spatially},capturing RNA in tissues and sequencing it to obtain a global map of gene expression. However, both approaches have their respective drawbacks.Imaging-based approaches offer higher spatial resolution, suitable for observing cellular and subcellular details. But they may have limitations in detecting the number of genes. NGS-based approaches provide a comprehensive understanding of global gene expression in tissues and can simultaneously detect a large number of genes, suitable for comprehensive bioinformatics analysis \cite{li2024stmcdi}. But they face two major challenges \cite{liu2022analysis}:1.lower spatial resolution. Such as 10X Visium \cite{rao2020bridging}, a spot with a diameter of 55 $\mu $m may contain 1 to 30 cells. In ST \cite{thrane2018spatially}, spot is 100 $\mu $m in diameter and may contain hundreds of cells. This makes it challenging to study individual differences between cells, such as cell subtypes, mutations, and expression heterogeneity. 2.Gaps Between Spots. The center-to-center distance between spots in ST is 200 $\mu $m, and in Visium, it is 100 $\mu $m. Clearly, there are significant uncovered spatial between spots, leaving large areas in the tissue unmeasured. This can affect researchers' comprehensive understanding of the entire cell population, hindering a full representation of the true state of the tissue, especially in scenarios where studying cell heterogeneity and local differences is crucial.

Here, we propose stEnTrans, a deep learning method using self-supervised learning \cite{jing2020self} that enhances the resolution of gene expression through prediction in unmeasured areas between spots to obtain a high-resolution and high quality gene expression profiles. stEnTrans requires no additional data, such as histology images, relying solely on spatial gene expression data as input.It comprehensively integrates the relevant information across different positions within genes and the correlations among different genes within the tissue. We applied stEnTrans to six datasets, including four real datasets sourced from the ST and 10X Visium platforms and two standard array datasets were simulated from STARmap and Stereo-seq data using coordinate mapping. The experimental results demonstrate the superiority of stEnTrans compared to other methods in terms of prediction accuracy and clarity of gene expression profiles, while also manifesting more biologically meaningful pathways.

\section{Method}

\begin{figure*}
    \centering
    \includegraphics[width=1\textwidth]{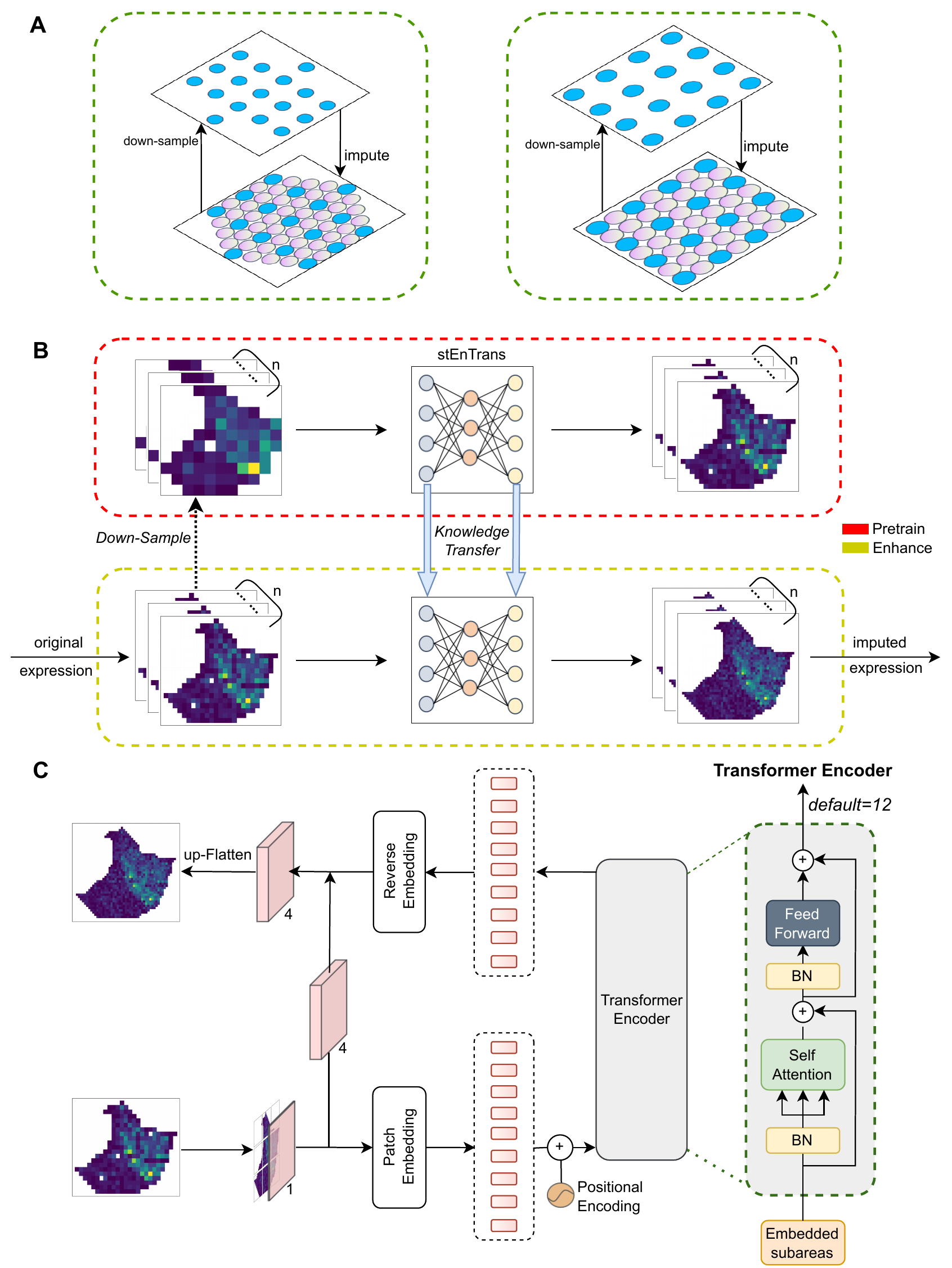}
    \caption{The network architecture of stEnTrans. (A) The schematic diagrams of down-sampling and imputation on honeycomb-based (left) and matrix-based (right) data. Down-sample: From left to right and top to bottom, alternately extract every spots and Extracted spots are non-adjacent to each other in the original gene expression profiles. Impute: Interpolate between adjacent spots. There are three types of adjacency: left-right, top-bottom and diagonal adjacency. (B) The self-supervised learning process of a network architecture. Pretraining the stEnTrans using downsampled gene expression as input and original gene pression as labels. After training the stEnTrans, inputting the original gene expression to obtain the high-resolution gene expression. (C) Details of network architecture. The trainable components mainly consist of Transformer Encoder modules.}
    \label{f1}
\end{figure*}

Our model undergoes two distinct phases: Pretrain and Enhance (Fig. \ref{f1}). During the Pretrain phase, We utilize down-sampling on the original gene expression profiles to generate low-resolution (LR) gene expression profiles, employing it as inputs, with the original gene expression profiles serving as the labels, and the outputs possesses the same resolution as the original gene expression profiles. After Pretrain phase, proceed to the Enhance phase, Using the original spatial gene expression profiles as inputs, high-resolution (HR) data is obtained. Our method is primarily designed for standard array data, focusing on ST and 10X Visium platforms. Additionally, for other platforms like STARmap and Stereo-seq, we can simulate them as array data through spatial coordinate mapping to achieve the purpose of data enhancement.

\subsection{Data Pre-processing}

stEnTrans only requires a gene expression matrix with spatial coordinates. We define the ST data as two matrices: (1) gene expression matrix,$\mathit{X}_{m\times n}$, which contains $\mathit{m}$ spots and $\mathit{n}$ genes.The value of $\mathit{X}_{ij} $ represents the expression level of $\mathit{j}$-th gene in the $\mathit{i}$-th spot. (2) spatial coordinate matrix, $\mathit{C}_{m\times 2}$. Its $\mathit{i}$-th row corresponds to the same spot as the $\mathit{i}$-th row of matrix $\mathit{X}_{m\times n}$. The values of $\mathit{C}_{i0}$ and $\mathit{C}_{i1}$ constitute the coordinate information of the $\mathit{i}$-th spot in two-dimensional space. We treat the expression of each gene in space as a sample.

\subsubsection{Data source}
All datasets analyzed in this paper are existing and publicly available. The human melanoma ST data (HM) can be found at \url{https://www.spatialresearch.org/resources-published-datasets/ doi-10-1158-0008-5472-can-18-0747} \cite{thrane2018spatially}. The STARmap mouse placenta (MP) can be found at \url{https://codeocean.com/capsule/9820099/tree/v1} \cite{he2021clustermap}. The Stereo-seq data from the adult mouse hemi-brain (AMHB) can be found \url{https://db.cngb.org/stomics/mosta/} \cite{chen2022spatiotemporal}. The human breast cancer spatial transcriptomics data (HBC) is available from the \url{https://www.10xgenomics.com/products/ xenium-in-situ/preview-dataset-human-breast} \cite{janesick2022high}. The Human Invasive ductal carcinoma spatial transcriptomics data (IDC) is available from the \url{https://support.10xgenomics.com/spatialgene-expression/datasets/1.2.0/V1_ Human_Invasive_Ductal_Carcinoma}. The mouse brain sagittal posterior data (MBSP) is available from the \url{https://www.10xgenomics.com/datasets/mouse-brain-serial-section-1-sagittal-posterior-1-standard-1-1-0}.

\subsubsection{Pre-processing for gene expression matrix}
The $\mathit{j}$-th column of $\mathit{X}$ represents the expression level of the $\mathit{j}$-th gene in the tissue. We denote this m-dimensional vector as $\mathit{X}^{j}$. Based on spatial coordinate information of all spots, $\mathit{X}^{j}$ can be mapped onto a plane and transformed into a matrix of shape ($\mathit{u}$, $\mathit{v}$). We denote this matrix as $\mathit{G}^{j}$, which can be considered as a gene expression profile. Specifically, the value at the $\mathit{r}_{i}$-th row and $\mathit{c}_{j}$-th column of $\mathit{G}^{j}$ represents the expression level of the $\mathit{j}$-th gene in the spot with coordinates ($\mathit{r}_{i}$, $\mathit{c}_{j}$).In addition, we need whether-in-tissue matrix $\mathit{M}$, which has the same shape as $\mathit{G}^{j}$, to indicate the presence or absence of spots. All its elements are either 0 or 1. For example, if $\mathit{M}_{\mathit{r}_{i}\mathit{c}_{j}}$ = 0, it indicates that there is not a spot with spatial coordinates ($\mathit{r}_{i}$, $\mathit{c}_{j}$); if $\mathit{M}_{\mathit{r}_{i}\mathit{c}_{j}}$ = 1, there is a spot.We use function R(·) to transform gene expression matrix with spatial coordinates into profile:

% 另一种表达形式：0\le j\le n-1, j \in \mathbb{N}_0 ---与DIST区分
\begin{equation}
    G^{j}=R(X^{j}, C), j=0, \cdots, n-1,
\end{equation}
where all profiles are denoted as the profile set $\mathit{G} = \left \{ G^{0}, \cdots ,  G^{n-1}\right \}$.

We use function T(·) to obtain in whether-in-tissue matrix from spatial coordinate matrix:
\begin{equation}
    M=T(C).
\end{equation}
where $M$ is the whether-in-tissue matrix corresponding to $G^{j}$.

Due to irregular tissue shapes, there may be spots outside the tissue in the gene expression maps. Additionally, due to limitations in current technology, some spots within the tissue may be unintentionally lost. We uniformly fill these occurrences with zeros in $G^{j}$.

\subsubsection{Data downsampling}
In the Pretrain phase, we need to create LR gene expression profiles through down-sampling (Fig.~\ref{f1}) after obtaining the original gene expression profiles. We assume that $G^{j}$ has $\mathit{u}$ rows and $\mathit{v}$ columns. This process can be described as:
\begin{equation}
\begin{gathered}
    G^{j}_{lr}(r_{i}, c_{j})=G^{j}(2r_{i}, 2c_{j}), \\
    r_{i}=0,1,\cdots,\lfloor \frac{u}{2} \rfloor -1, c_{j}=0,1,\cdots,\lfloor \frac{v}{2} \rfloor -1, j=0,1,\cdots,n-1,
\end{gathered}
\end{equation}
where the shape of $G^{j}_{lr}$ is ($\lfloor \frac{u}{2} \rfloor$, $\lfloor \frac{v}{2} \rfloor$). We use function $D(\cdot)$ to summarize the above process:
\begin{equation}
    G^{j}_{lr}=D(G^{j}), j=0,1,\cdots,n-1,
\end{equation}
where all profiles $G^{j}_{lr}$ are denoted as the LR profile set $\mathit{G}_{lr} = \left \{ G^{0}_{lr}, G^{1}_{lr}, \cdots ,  G^{n-1}_{lr}\right \}$.

\subsection{Self-supervised learning}

In the realm of data enhancement for spatial transcriptomics, where no labels are available, so we employ a self-supervised learning approach \cite{jing2020self}. We design auxiliary tasks that utilize the data itself as supervisory information. The entire process is divided into two phases, denoted as Pretrain and Enhance (Fig. \ref{f1}). In Pretrain phase, we train the model using both LR and original gene expression profiles. Then in Enhance phase, we input the original data and obtain HR gene expression profiles.

\subsubsection{Pretrain phase}

stEnTrans achieves gene expression interpolation by employing Transformer Encoder \cite{vaswani2017attention} to extract global features of the spatial distribution of all genes within the tissue, thereby extending the size of the gene expression profiles from ($\mathit{u}$, $\mathit{v}$) to (2$\mathit{u}$, 2$\mathit{v}$) (Fig. \ref{f1}). In this phase, we train model on the LR gene set $\mathit{G}_{lr} = \left \{ G^{0}_{lr}, G^{1}_{lr}, \cdots ,  G^{n-1}_{lr}\right \}$ and original gene set $\mathit{G} = \left \{ G^{0}, G^{1}, \cdots ,  G^{n-1}\right \}$. original gene set $\mathit{G} $ is constructed from gene expression matrix $\mathit{X}_{m \times n}$ and coordinate matrix $\mathit{C}_{m \times 2}$ by function $R(\cdot )$. LR gene set $\mathit{G}_{lr}$ is constructed from original gene set $\mathit{G} $ by function $D(\cdot )$. We use $\Phi (\cdot )$ represent the network model, which takes input $G_{lr}$ and aims to recover $G$. After filtering through  whether-in-tissue matrix $\mathit{M}$, we denote the output as $\mathit{G}^{\prime }$, which has the same shape as $G$. We use $\theta $ represent the parameters of the network model. Finally, we use Mean Squared Error to train our network model. The above process can be described as:

\begin{equation}
    G^{\prime }=\Phi (G_{lr}; \theta ) \cdot M,
\end{equation}
\begin{equation}
    Loss(\theta )=\frac{1}{n} \sum_{j=0}^{n-1} \left \| {G^{j}}^{\prime } - G^{j} \right \| ^{2}.
\end{equation}

\subsubsection{Enhance phase}

stEnTrans can adapt well to inputs of different sizes. Therefore, after the Pretrain phase, we can enhance original data to HR gene expression profiles using stEnTrans:

\begin{equation}
    {G_{hr}}=\Phi (G; \theta ) \cdot H(M),
\end{equation}
where $G_{hr}$, $H(M)$ have shapes $(n, 2u, 2v)$ and $(2u, 2v)$ respectively, $H(\cdot )$ can learn the boundary on the imputed gene expression by $\mathit{M}$ to generate a new whether-in-tissue matrix corresponding to $G_{hr}$. if location $(c_{i}, r_{j})$ is the initial spot or lies between the initial two spots, then $H(M)(c_{i}, r_{j})=1$, otherwise, it equals 0 (Fig.~\ref{f1}).

Finally, we can get their corresponding gene expression matrix $X_{hr}$ and spatial coordinate matrix $C_{hr}$ from $G_{hr}$ and $H(M)$ by applying the inverse functions of $R(\cdot )$ .

\subsection{Details of the proposed stEnTrans}

Transformer uses self-attention mechanisms to capture dependencies among different positions in the input sequence, modeling interactions among all tokens \cite{vaswani2017attention} (Fig. \ref{f1}). The standard Transformer architecture is designed for processing 1-D sequential data. To adapt it for handling 2-D gene expression maps, we reshape the maps into a series of patches \cite{dosovitskiy2020image}. $P$ represents the height and width of the patches. Suppose that size of the input $\mathit{x}$ is $(H, W)$, so the desired size of the output is $(2H, 2W)$. We achieve this by applying four trainable convolutional structures, where the convolutional kernel size and stride are both set to patch size, followed by concat, and finally flattening to map the tokens to $4\times P^{2}$ dimensions. In addition, if the size of the input profile are not multiples of the patch size, we need to perform zero-padding on the profile. This process corresponds to the patch embedding in stEnTrans:
\begin{equation}
    x_{k} = x\ast W_{k} + b_{k}, k=0,1,2,3,
\end{equation}
\begin{equation}
    x^{\prime } = \text{concat}(x_{0},x_{1},x_{2},x_{3}),
\end{equation}
\begin{equation}
    \hat{x} = \text{Flatten}(x^{\prime }),
\end{equation}
where $\mathit{x}$, $x_{k}$, $x^{\prime }$, $\hat{x}$ have shapes $(H, W)$, $(P^{2}, H/P, W/P)$, $(4, P^{2}, H/P, W/P)$ and $(N, 4P^{2})$ respectively, $W_{k}$ and $b_{k}$ are the filter parameters and biases of convolutional structures. Here, $N=HW/P^{2}$, which represents the number of generated patches and is also the input sequence length for the Transformer Encoder. We refer to above process as Patch Embedding.

Due to the fact that the self-attention mechanism itself does not inherently contain information about the position of tokens in the sequence, it is necessary to introduce positional encoding to help the model understand the order of the input sequence. The relative positional embeddings as described in \cite{shaw2018self} is not applicable to our method as the sequence lengths differ between phases Pretrain and Enhance. so stEnTrans adopts sine and cosine functions with different frequencies to encode positional information:
\begin{equation}
\begin{gathered} 
    Pos(\hat{x} )(p,q)= \begin{cases}
    \sin(p/10000^{2q/d }), & p \bmod 2 = 0 \\
    \cos(p/10000^{2(q-1)/d } ), & p \bmod 2 = 1
    \end{cases}\\
    p=0,1,\cdots, N-1, q=0,1,\cdots, 4P^{2},
\end{gathered}
\end{equation}
\begin{equation}
    \hat{x}_{0} = \hat{x} +Pos(\hat{x}),
\end{equation}
where $Pos(\hat{x})$ and $\hat{x}$ have the same shape $(N, 4P^{2})$, $p$ is the position, $q$ is the dimension and $d$ is the patch embedding dimension.

Then, stEnTrans extracts global features of the LR profiles through the Transformer Encoder, which consists of multiple Transformer blocks. The Transformer block is mainly  composed of multiheaded self-attention (MSA) and MLP blocks. LayerNorm is applied before each block and residual connections are applied after each block \cite{brody2023expressivity,liu2021rethinking}.

Assuming the input of the MSA block is $U$ and the output is $Z$, here is a description of the multi-head self-attention block:
\begin{equation}
    [Q_{i}, K_{i}, V_{i}] = UW_{i} + b_{i}, i = 1\dots h,
\end{equation}
\begin{equation}
    head_{i}=\text{softmax}(\frac{Q_{i}K^{T}_{i}}{\sqrt{d_{i}}})V_{i}, i = 1\dots h,
\end{equation}
\begin{equation}
    Z=\text{concat}(head_{1},\dots ,head_{h}),
\end{equation}
where $h$ denotes the numbers of different self-attention operations in multi-head self-attention, The queries, keys, and values are linearly projected $h$ times with different learnable parameters, $d$ is set to a fixed shape of $4P^{2}/h$ and $W_{i}$ have the shape of $(4P^{2}, 4P^{2}/h)$. This parallel processing allows the model to capture diverse patterns and relationships. We can represent the above process using the function MSA$(\cdot)$:
\begin{equation}
    Z=\text{MSA}(U).
\end{equation}

In the MLP sub-layer, there are two linear transformations with a GELU activation function in between. Suppose that the input is $Z$, the process can be described as follows:
\begin{equation}
    \text{MLP}(Z)=\text{GeLu}(ZW_{1}+b_{1})W_{2}+b_{2},
\end{equation}
where $Z$ and $\text{MLP}(Z)$ hava the same shape $(N, 4P^{2})$. Assuming $D=4P^{2}$, then $W_{1}\in \mathbb{R} ^{D\times 2D}$ and $W_{2}\in \mathbb{R} ^{2D\times D}$.

Assuming the input of the $\ell$-th Transformer block is $\hat{x}_{\ell-1}$, we denote its output as $\hat{x}_{\ell}$, which is also the input of the $\mathit{\ell\!+}$\!1-th block. Specifically, $\hat{x}_{0}$ serves as the input to the first block. The entire process of gene expression maps enhancement in stEnTrans within the Transformer Encoder can be described as follows:
\begin{align}
    & \hat{x}^{\prime }_{\ell} = \text{MSA}(\text{LN}({\hat{x}_{\ell-1}})) + \hat{x}_{\ell-1}, & \ell=1,\dots, N, \\
    & \hat{x}_{\ell} = \text{MLP}(\text{LN}(\hat{x}^{\prime }_{\ell})) + \hat{x}^{\prime }_{\ell}, & \ell=1,\dots, N,
\end{align}

After passing through the Transformer Encoder,  we perform a reverse embedding operation on the 1D sequence of token embeddings $\hat{x}_{N}$, transforming it into 4-channel HR gene expression profile:
\begin{equation}
    y^{\prime} = \text{Flatten}^{-1}(\hat{x}_{N}),
\end{equation}
\begin{equation}
\begin{gathered}
    \Delta y(c,i,j)=y^{\prime}(c,r,\left \lfloor \frac{i}{P}  \right \rfloor,\left \lfloor \frac{j}{P}  \right \rfloor ), c=0,1,2,3,\\
    r=P(i \bmod P)+(j \bmod P), i=0,\dots,H\!-\!1,j=0,\dots,W\!-\!1 ,
\end{gathered}
\end{equation}
where $y^{\prime}$, $\Delta y$ have shapes $(4, P^{2}, \frac{H}{P}, \frac{W}{P})$ and  $(4, H, W)$ respectively, $r$ represents the position within a patch and its value increments from left to right and top to bottom within each patch, starting from 0.

Next, inspired by residual networks \cite{he2016deep}, we directly combines the original data and each channel of $\Delta y$. In this case, we set $\Delta y=[\Delta y_{0}, \Delta y_{1}, \Delta y_{2}, \Delta y_{3}]$, where $\Delta y_{c}$ hava the same shape $(H, W)$, we get:
\begin{equation}
    y = [y_{0}, y_{1}, y_{2}, y_{3}] = [\Delta y_{0}+x, \Delta y_{1}+x, \Delta y_{2}+x, \Delta y_{3}+x].
\end{equation}

Finally, we can obtain HR gene expression profile by merging the 4-channel expression profile $Y$. We denote the HR expression profile as $y_{hr}$, and the calculation process is as follows:
\begin{equation}
    y_{\text{\tiny{hr}}}(p,q) = \begin{cases}
    y(0, \left \lfloor \frac{p}{2} \right \rfloor , \left \lfloor \frac{q}{2} \right \rfloor ), & p \bmod 2 = 0, q \bmod 2 = 0 \\
    y(1, \left \lfloor \frac{p}{2} \right \rfloor , \left \lfloor \frac{q}{2} \right \rfloor ), & p \bmod 2 = 0, q \bmod 2 = 1 \\
    y(2, \left \lfloor \frac{p}{2} \right \rfloor , \left \lfloor \frac{q}{2} \right \rfloor ), & p \bmod 2 = 1, q \bmod 2 = 0 \\
    y(3, \left \lfloor \frac{p}{2} \right \rfloor , \left \lfloor \frac{q}{2} \right \rfloor ), & p \bmod 2 = 1, q \bmod 2 = 1
    \end{cases},
\end{equation}
where $y_{\text{\tiny{hr}}}$ is divided into a series of non-overlapping 2x2 sub-maps, each sub-map's top-left, top-right, bottom-left, and bottom-right respectively come from channels 0, 1, 2, and 3 of $y$.

\section{Experimental Results}

\begin{figure*}
    \centering
    \includegraphics[width=1\textwidth]{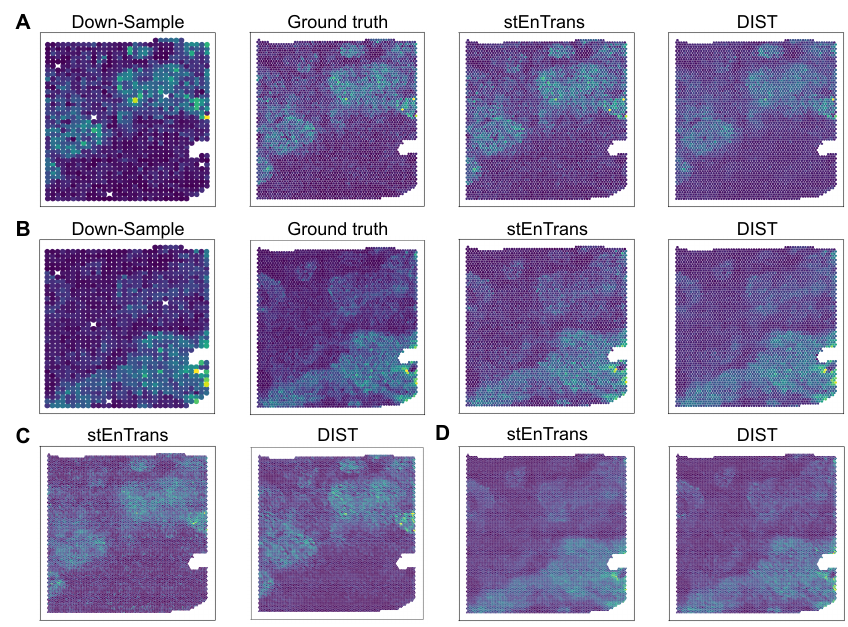}
    \caption{The experiments indicate that stEnTrans exhibits superior interpolation performance. (A) and (B) Applying stEnTrans, DIST, NEDI, Linear, Cubic and NN to downsampled simulated data for predicting the ground truth of gene expression. This gene ZNF703 comes from 10X Visium IDC data. (B) A series of gene expression profiles are obtained through a consistent process with (A). This gene MUC1 also originates from 10X Visium IDC data. (C) and (D) apply stEnTrans and DIST to the ground truth of ZNF703 gene and MUC1 gene, resulting in HR gene expression maps. Due to inferior interpolation performance of other methods, we only compare with DIST. The gene-wise minimum and maximum of the ground truth map the color ranges consistently.}
    \label{f2}
\end{figure*}

\subsection{More finer interpolation capability}
To validate the interpolation capability of stEnTrans, we evaluated two genes with different spatial patterns on IDC data, ZNF703 and MUC1. We used down-sampled data as simulation, the original data as ground truth, and applied stEnTrans, DIST \cite{zhao2023dist}, Linear, Cubic, NN and NEDI \cite{li2001new} for interpolation on the down-sampled data for predicting ground truth (Fig. \ref{f2}). In real-world applications, gene expression exhibits vacant locations due to imperfect spatial coverage and quality control, so down-sampled data includes losses to mimic these real-world scenarios. stEnTrans and the mentioned interpolation methods can all fill vacancy expression to ensure the integrity and continuity of tissue slices. The result shows that stEnTrans demonstrated a more delicate effect in interpolation, closely resembling real data than other methods. NN and NEDI may result in discontinuous expression with noticeable aliasing. Linear and Cubic methods overcome aliasing but result in blurring. These interpolation methods only rely on nearby points to predict the gene expression of unmeasured points, neglecting the utilization of global effective information. Furthermore, stEnTrans is trained on all genes within the tissue, effectively establishing correlated information among genes, as opposed to methods that only consider the expression of individual gene. DIST enhances linear continuity but lacks clarity in details, deviating from ground truth. For modeling long-range dependencies, convolutional methods lack the robustness of the Transformer framework, making stEnTrans superior to DIST in this aspect.

Next, we applied stEnTrans and DIST to genes ZNF703 and MUC1 for interpolation on the ground data to obtain their HR gene expression profiles (Fig. \ref{f2}). Through comparison, while both methods can achieve HR data with the same resolution, stEnTrans presents a smoother expression mapping, sharper gene expression, and more detailed depiction.

\subsection{More accurate interpolation capability}

To further illustrate the superiority of stEnTrans, we calculated the Pearson correlation coefficients (PCC) between ground truth and each imputed expression for all spots of the genes after quality filtering across six datasets.

\begin{figure*}[t]
    \centering
    \includegraphics[width=1\textwidth]{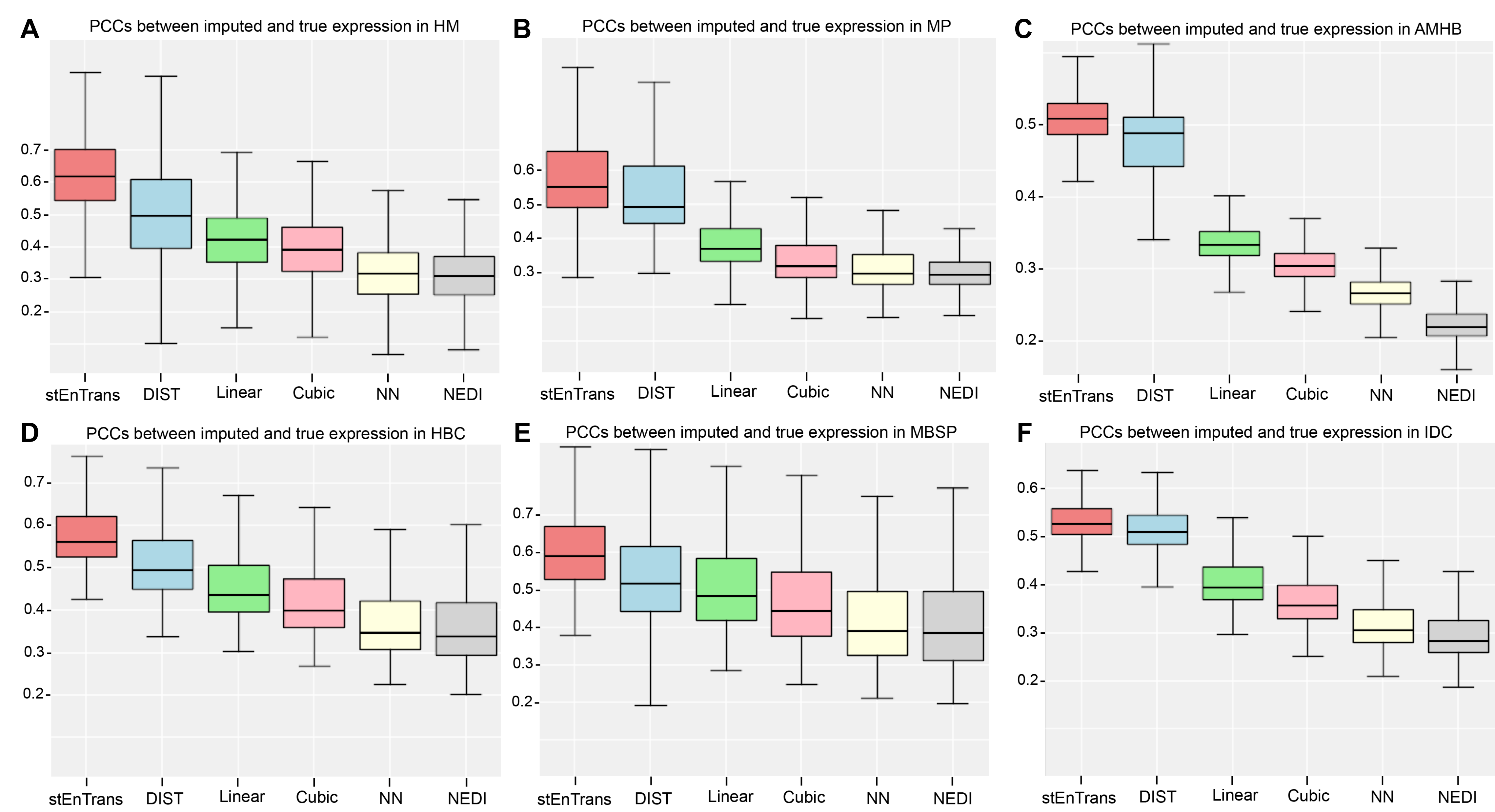}
    \caption{PCCs evaluation between ground truth and each imputed expression in six ST datasets. The results indicate that stEnTrans exhibits outstanding precision in interpolation, as demonstrated through experiments on six datasets. (A) The human melanoma ST data (mel1 rep1). (B) Simulation created from STARmap mouse placenta. (C) Simulation created from Stereo-seq adult mouse hemi-brain. (D) Human breast cancer spatial transcriptomics data. (E) Visium mouse sagittal posterior brain. (F) Visium human IDC data. Gene-wise Pearson correlation correlations between ground truth and imputed expression using stEnTrans, DIST, Linear, Cubic, NN, NEDI on that six datasets. Boxes represent the middle 50\% range of the data,which show 25th, 50th and 75th percentiles. Whiskers represent the overall distribution range of the data, providing information about the data's dispersion and potential outliers.}
    \label{f3}
\end{figure*}

To illustrate that stEnTrans can also be applied to other platforms, we created a simulated ST data from STARmap data. By proportionally mapping the coordinates of these pseudo-spots in spatial dimensions and fine-tuning, we can simulate array-based ST data. RNA molecules of 903 genes have been determined in the mouse placenta for this dataset. ClusterMap clusters RNA into subcellular structures, generating 7224 pseudo-spots representing subcellular sizes \cite{he2021clustermap}, which closely align with our simulated data. Continuing, we simulated another ST data from Stereo-seq data, which provides subcellular resolution spatial expression in the adult mouse hemi-brain \cite{chen2022spatiotemporal}, using the same spatial coordinate mapping method as described above. Additionally, there is one real ST dataset and three 10X Visium datasets. We calculated PCCs between ground truth and each imputed expression for all spots of the genes after quality filtering across the aforementioned six datasets and showed the results in (Fig. \ref{f3}).

stEnTrans achieved higher median in gene prediction PCCs between imputed and true expression compared to all other methods across six different datasets (median: in HM data, stEnTrans=0.636; in MP data, stEnTrans=0.554; in AMHB data, stEnTrans=0.513; in HBC data, stEnTrans=0.584; in MBSP data, stEnTrans=0.606; in IDC data, stEnTrans=0.537). Even in HM data, HBC data and MP data, the 25th percentile of PCCs for stEnTrans surpasses or is nearly equivalent to the 75th percentile for DIST. In the other three datasets, stEnTrans’s 25th percentile of PCCs are higher than the maximum of 50th percentiles of DIST. We did not compare stEnTrans with methods like BayesSpace \cite{zhao2021spatial}, TESLA \cite{hu2023deciphering} and iStar \cite{zhang2024inferring} because BayesSpace divides a spot into multiple sub-spots, rendering it unable to predict unmeasured regions between spots; TESLA and iStar requires histological images, which some simulated data cannot provide.

\begin{figure*}
    \centering
    \includegraphics[width=1\textwidth]{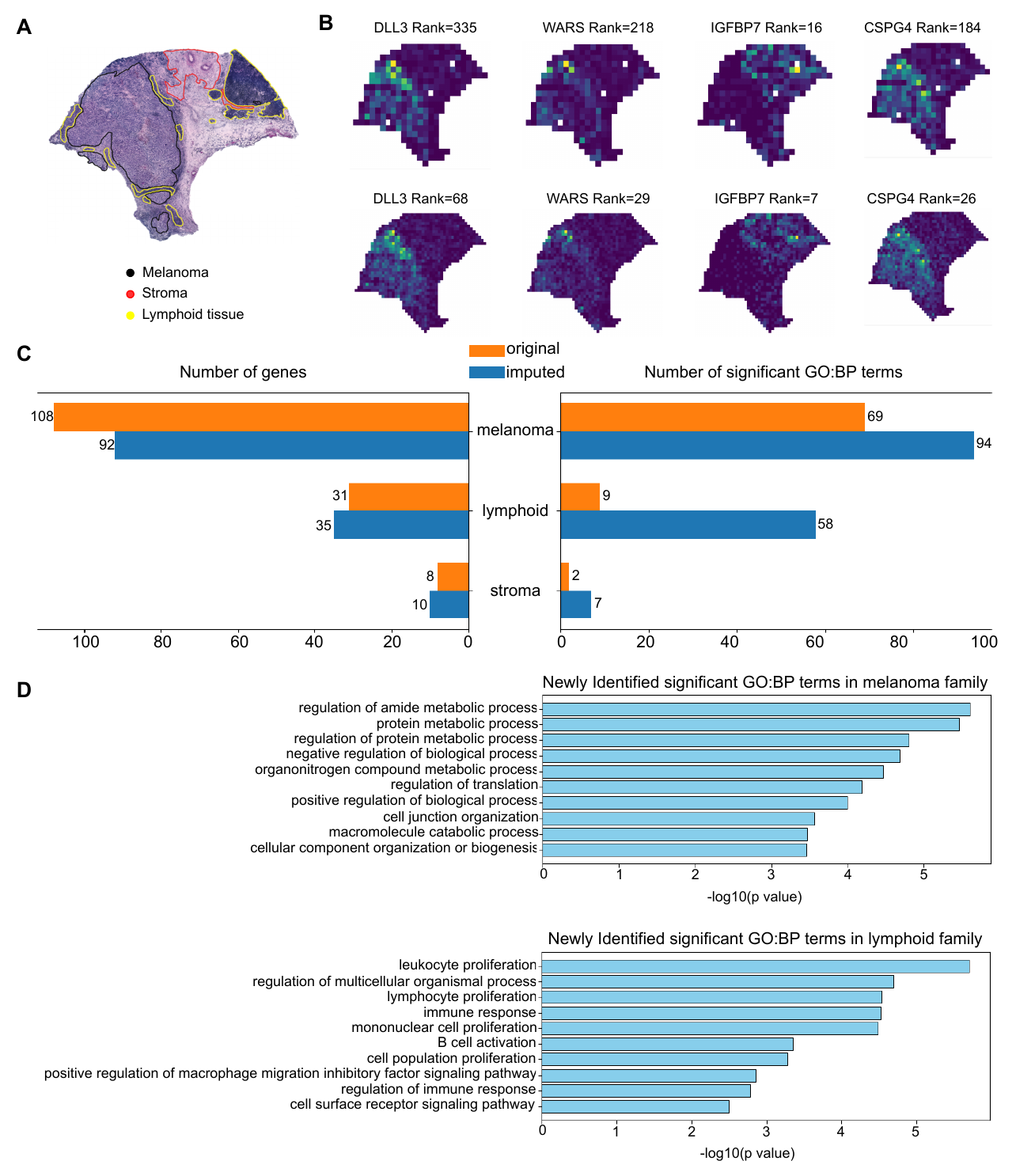}
    \caption{stEnTrans can help find spatial patterns for ST data and enrich more biologically significant pathways. (A) 
The original histopathological annotations of hematoxylin- and eosin-stained tissue image, where black represents melanoma, red represents stroma, and yellow represents lymphatic tissue. (B) stEnTrans has endowed disease-related genes with distinct spatial patterns. These genes exhibit higher rankings in the imputed data, whereas they have lower rankings in the original data. We provided the rankings of genes at the top of each gene expression profile. (C) The number of genes and significant GO:BP terms in three mainly families of original data and imputed data, where orange represents original data and blue represents imputed data. (D) Top 10 significant GO:BP terms were only identified in the imputed data. above corresponds to the melanoma family, and below corresponds to the lymphoid family.}
    \label{f4}
\end{figure*}

\subsection{Better help to discover spatial patterns}

Next, we explore whether stEnTrans can help discover spatial patterns in melanoma ST data. We utilize a tool, Sepal, which employs diffusion-based modeling to identify transcripts with spatial patterns in the transcript profiles \cite{andersson2021sepal}. This method is applied to original and imputed mel ST data using stEnTrans. Following Sepal's standards, we rank each transcript profile according to their degree of randomness, and then extract the top-150 transcript profiles as experimental samples. Next, we categorize the top-150 transcript profiles into pattern families based on the similarity of their spatial organization. Among them, the spatial organization structures and biological processes of the three main pattern families can match well with histological annotations, including melanoma, lymphoid, and stroma. Finally, we perform functional enrichment analysis for each pattern family using the Gene Ontology: Biological Processes (GO: BP) database \cite{raudvere2019g}.

We selected a subset of transcript profiles that ranked lower in the original melanoma ST data but higher in the imputed melanoma ST data (Fig.~\ref{f4}). These genes play important roles in various physiological and pathological processes. DLL3 is highly expressed in melanoma, making it a potential therapeutic target \cite{yao2022dll3}. WARS plays important physiopathological roles in cancer diseases, with the potential to serve as a pharmacological target and therapeutic agent \cite{ahn2021tryptophanyl}. The expression of IGFBP7 can induce cellular senescence and apoptosis, and it has the potential to inhibit the growth of melanoma \cite{chen2010vivo}. Due to low resolution, incomplete gene expression results in these genes being ranked lower in the original melanoma data. After enhancing the resolution through stEnTrans, the spatial structure of these genes becomes more apparent, resulting in a significant improvement in their rankings.

We first separately counted the respective quantities of the main three pattern families within the top-150 genes in original and imputed human melanoma data and then performed functional enrichment analysis (Fig. ~\ref{f4}). The bar chart reveals that, despite a decrease in the number of genes in the melanoma family, it has also enriched more GO:BP terms. the number of genes and GO:BP terms in the lymphoid and stroma families have both increased. It's worth noting that the imputed human melanoma data for the three families collectively enriched in 79 new pathways compared to the original human melanoma data. In the melanoma family, the top-10 significantly enriched new pathways are mostly related to biological regulation and metabolic processes. In contrast, within the lymphoid family, the top-10 significantly enriched new pathways are mostly associated with immune response, cell activation, and proliferation processes. In addition to the top-10 significantly enriched new pathways, the imputed human melanoma data also newly enriched in multiple pathways related to growth factors and lymphocyte activation.

\subsection{Ablation Study}
% simulated ST data from STARmap mouse placenta (MP), simulated ST data from Stereo-seq adult mouse hemi-brain (AMHB), human melanoma ST data (HM), Human Invasive ductal carcinoma spatial transcriptomics data (IDC), human breast cancer spatial transcriptomics data (HBC), mouse brain sagittal posterior data (MBSP)
To assess the contributions of absolute positional encoding (denote as Pos) and the Res module to performance, we conducted a ablation study. Our ablation study is based on simulated ST data from STARmap mouse placenta (MP), simulated ST data from Stereo-seq adult mouse hemi-brain (AMHB), human melanoma ST data (HM), Human Invasive ductal carcinoma spatial transcriptomics data (IDC), human breast cancer spatial transcriptomics data (HBC), mouse brain sagittal posterior data (MBSP). All experiments are evaluated using PCC.

Specifically, we initially assessed the performance of the original model by calculating PCCs on six datasets as the baseline for comparison. Subsequently, we systematically removed Module Pos, Module Res, and both simultaneously, observing the model's performance under these ablation conditions. Overall, the results indicate that, among all ablation conditions, the removal of Pos has the most significant impact on the model's performance, resulting in the poorest performance. Next in significance is the simultaneous removal of Pos and Res, while the impact of removing Res is relatively smaller but still leads to a decline in performance.

\begin{table*}[!h]
\caption{Effects of Pos and Res. ``Pos'' refers to the absolute positional encoding, while ``Res'' stands for the skip connection that combines the original data to the output of the Transformer Encoder.}
\label{t1}
\begin{tabular*}{\hsize}{@{}@{\extracolsep{\fill}}cccccccc@{}}
\hline
\multicolumn{2}{c}{\textbf{PCCs}} & HM    & MP    & AMHB    & HBC   & MBSP   & IDC           \\ 
\hline
\multicolumn{2}{c}{stEnTrans}         & \textbf{0.636}\tiny{$\pm$0.018} & \textbf{0.554}\tiny{$\pm$0.021} & \textbf{0.513}\tiny{$\pm$0.056} &  \textbf{0.584}\tiny{$\pm$0.006} &
\textbf{0.606}\tiny{$\pm$0.010} & \textbf{0.537}\tiny{$\pm$0.003} \\
\multicolumn{2}{c}{(w/o)Pos}          & 0.581\tiny{$\pm$0.017} & 0.439\tiny{$\pm$0.019} & 0.506\tiny{$\pm$0.005} & 0.559\tiny{$\pm$0.009} & 0.569\tiny{$\pm$0.016} & 0.525\tiny{$\pm$0.004} \\
\multicolumn{2}{c}{(w/o)Res}          & 0.619\tiny{$\pm$0.014} & 0.476\tiny{$\pm$0.020} & 0.510\tiny{$\pm$0.005} & 0.570\tiny{$\pm$0.008} & 0.598\tiny{$\pm$0.012} & 0.529\tiny{$\pm$0.004} \\ 
\multicolumn{2}{c}{(w/o)Pos\&Res}     & 0.587\tiny{$\pm$0.017} & 0.470\tiny{$\pm$0.015} & 0.505\tiny{$\pm$0.004} & 0.561\tiny{$\pm$0.009} & 0.583\tiny{$\pm$0.014} & 0.529\tiny{$\pm$0.004} \\ 
\hline
\end{tabular*}
\end{table*}

\section{Discussion And Conclusion}
In our research, we propose stEnTrans, a deep learning method based on Transformer, which can impute the gene expression on unmeasured areas that are not covered between spots and accidentally lost locations during sequencing through self-supervised learning, enhance the expression of all spots, and thus comprehensively improve the quality of spatial transcriptomics data.Our method does not rely on any other data, such as histological images, and only requires a gene expression matrix with spatial coordinates to improve gene expression profiles to high resolution. Compared with other methods of similar function, it has higher precision and the obtaining high-resolution gene expression profiles have more delicate and smooth characteristics.

In spatial transcriptomics studies, identifying genes with spatial expression patterns is very important for researching various physiological and pathological processes. 
However, due to limitations in existing technologies, the statistical power is often low, making it challenging to identify genes with spatial expression patterns within small local regions using raw spatial transcriptomic data. stEnTrans has the capability to enhance the resolution of gene expression profiles, which can help to discover the spatial patterns of genes, making disease-related genes have more significant spatial patterns. In addition, it can also help to discover more biologically meaningful pathways.

The results of ablation experiments show that both Res module and absolute position encoding are indispensable to the network and can improve the performance of the network, but the use of absolute position encoding has a higher impact on improving the network's performance.

% 接下来的研究方向是：在不依赖其它任何数据的前提下，不仅可以通过对未测量区域插值来提高分辨率，还可以通过将spot划分为sub-spots,将分辨率提高到子点水平
% BayesSpace:不依赖其它数据，但只能将spot划分为sub-spots，不能对未测量区域预测
% TESLA(李明瑶):超分，可以对未测量区域预测，也可以将spot划分为sub-spots，但是需要依赖组织学图像
% stEnTrans & DIST:不依赖组织学图像，可以对未测量区域预测，但无法将spot划分为sub-spots(即无法缩小spot面积)
Despite the significant achievements of stEnTrans in improving the resolution of gene expression profiles by imputing, there are still several challenges in practical applications. Our next research direction is not only to impute the unmeasured areas, but also to increase the resolution to the sub-spot level by dividing the spot into multiple sub-spots without relying on any other data.

\subsubsection{Acknowledgements} 
The work was supported in part by the National Natural Science Foundation of China (62262069), in part by the Yunnan Fundamental Research Projects under Grant (202201AT070469, 202301BF070001-019).
%in part by the Program of Yunnan Key Laboratory of Intelligent Systems and Computing (202205AG070003). 


\bibliographystyle{splncs04}
\bibliography{isbra}
\end{document}